\documentstyle[12pt]{article}

\def\a{\alpha}
\def\o{\Omega}
\def\p{\phi}
\def\P{\Phi}
\def\cg{\cal G}
\def\cb{\cal B}
\def\pr{\partial }
\def\ca{\cal A}
\newcommand{\be}{\begin{equation}} \newcommand{\ee}{\end{equation}}
\newcommand{\bea}{\begin{eqnarray}}\newcommand{\eea}{\end{eqnarray}}

\begin{document}
\baselineskip= 24 truept
\begin{titlepage}

\title { Target Space Structure of a Chiral Gauged Wess-Zumino-Witten
Model}


\author{Supriya K. Kar$^1$ and Alok Kumar$^2$\\
Institute of Physics, Bhubaneswar-751005, INDIA.}


\footnotetext[1]{e-mail:supriya\%iopb@shakti.ernet.in}
\footnotetext[2]{e-mail:kumar\%iopb@shakti.ernet.in}

\date{}
\maketitle

\thispagestyle{empty}

\vskip 1in
\begin{abstract}

\vskip .3in

The background for string propagation is obtained by a chiral gauging
of the $SL(2,R)$ Wess-Zumino-Witten model. It is shown explicitly
that the resulting background fields satisfy the field equations of the
three dimensional string effective action and the target space has
curvature singularity. Close connection of our solution with the three
dimensional black string is demonstrated.

\end{abstract}

\bigskip
\vfil

\rightline{IP/BBSR/92-18}
\rightline{March '92.}
\end{titlepage}

\eject

The target space structure of the gauged Wess-Zumino-Witten
$(WZW)$ model has been investigated by Witten $[1]$ and others
$[2]$. It was shown $[1]$ that the vector or axial
gauging of a $U(1)$ subgroup
of the $SL(2,R)$ $WZW$ model leads to the string propagation in a
two dimensional target space with black hole singularity.
There have been several generalizations of this result. In particular,
two dimensional charged black hole $[3]$,
three dimensional black string $[4]$ as well as four dimensional
solutions with curvature singularities $[5,6]$ have been constucted
as gauged $WZW$ models. It has also been
demonstrated that the two gaugings mentioned above are related by duality
$[7,8]$ and they are anomaly free.

In a recent paper $[9]$, it has been shown that
there is another gauging for the two dimensional $WZW$ model, called "chiral
gauging",which is also anomaly free. Quantization  of the chiral $WZW$
model and their applications to the coset constructions have been done in
ref.$[9]$.

In this paper, we investigate the target space structure of the
$SL(2,R)$ $WZW$ model with a chiral $U(1)$ gauging. It is shown that
the resulting model describes a three dimesional target space with
curvature singularity. Furthermore, by an $O(2,2)$ transformation
acting on background metric $(G_{\mu\nu })$, dilaton $({\p })$ and
antisymmetric tensor $(B_{\mu\nu })$, it is demonstrated explicitly
that our solution can be transformed to the three dimensional uncharged
black string of ref.$[4]$. The connection to the charged black
string $[4]$ is also discussed.

We start by writing down the action for the chiral gauged
Wess-Zumino-Witten $(CGWZW)$ model $[9]$ :
\be
S = S^{WZW} + {k\over 2\pi }\int d^2z\; Tr\; [\; A^{R}_{z}\;
U^{-1}\bar{\pr }U + A^{L}_{\bar z}\; {\pr }U\;
U^{-1} + A^{R}_{z}U^{-1}A^{L}_{\bar z}U\; ]
\ee
where,
\bea
S^{WZW}=&&{k\over {4\pi }}\int d^2z\; tr\; [U^{-1}{\pr }U\;
U^{-1}\bar{\pr }U]\nonumber\\
&&+\; {k\over {12\pi }}\int d^2x\; dt\; {\epsilon }^{ijk}
Tr\; [U^{-1}{\pr }_iU\; U^{-1}{\pr }_jU\; U^{-1}{\pr }_kU ]
\eea
\noindent and $A^R_z\; (A^L_{\bar z})$ in eq.(1) is the $z (\bar
z)$ component of gauge field $A^R_{\mu }\; (A^L_{\mu })$.
The classical action $S$ is invariant under the gauge
transformation $[9]$,
\bea
&&\delta U = v_L\; (\bar z)\; U - U\; v_R(z),\nonumber\\
&&\delta A^{L}_{\bar z}= - \bar\pr v_L\; ({\bar z})\nonumber\\
and &&\delta A^{R}_{z} = \pr v_R\; ({z}).
\eea

\noindent It has been shown in ref.$[9]$ that the
theory is anomaly free. Therefore the gauge invariance is
maintained at the quantum level as well. For comparison, we also
write down the action for the vector $GWZW$ model $[1,10,11]$ :
\bea
S^V=&&S^{WZW} \nonumber\\&&+ {k\over 2\pi }\int d^2z\; Tr\; [\; A\;
U^{-1}{\bar\pr }U + \bar A\; {\pr }U\;
U^{-1} + A\;U^{-1}\bar A\; U + \bar A\;A\; ]
\eea
\noindent where in eq.(4) $A$ and $\bar A$ are the two light cone components
of a gauge field $A_{\mu }$.

 From eqs.(1) and (4), it is observed that, except for the absence
of one of the terms quadratic in gauge fields, the expression for the
$CGWZW$ model $[9]$ is similar to that of the vector (or axial) one
$[1,10,11]$.
However it will be seen that the absence of this term makes crucial
difference in the target space structure.

We now explicitly work out the case of the $\; SL(2,R)\; $ $WZW$
model when  "chiral" gauging is done with respect to the $U(1)$ subgroup
generated by the currents
$J^L_{\bar z} \equiv\; -{i\over 2}\; Tr\; ({\sigma }_2\;
U^{-1}{\pr }_{\bar z}U)
\;\; and \;\; J^R_{z} \equiv\; -{i\over 2}\; Tr\; (\sigma _2\;\pr _zU\;
 U^{-1}).$
\noindent Then by representing the gauge fields as, $A^L_{\bar z} \equiv
\big (-{i\over 2}{\ca }^L_{\bar z}\;\sigma _2\big )$, $A^R_z \equiv
\big (-{i\over 2}{\ca }^R_z\;\sigma _2\big )$ and the
sigma model field $U$ as,
\be
U\equiv exp\; {\big ({i\over 2}\phi _L\;\sigma _2\big )}\; exp\;
{\big ({r\over 2}\;\sigma _1\big )}\; exp\; {\big ({i\over 2}\phi _R\;
\sigma _2\big )}
\ee
\noindent one obtains $[10]$,
\be
S^{WZW} = {k\over 4\pi } \int d^2z\; (-{1\over 2}{\pr }{\phi
}_L{\bar\pr }{\phi }_L\; -\; {1\over 2}\pr {\phi
}_R{\bar\pr }{\phi }_R\nonumber\\
+\; {1\over 2}\pr r\bar\pr r -
cosh\; r\; \pr {\phi }_R \bar\pr {\phi }_L),
\ee
\noindent and
\bea
S =&&S^{WZW}\; +\; {k\over 4\pi } \int d^2z\; [{\ca }^R_z(\bar\pr
{\phi }_R\; +\; cosh\; r\;\; \bar\pr {\phi }_L)\nonumber\\
&&+\; {\ca }^L_{\bar
z}(\pr {\phi }_L\; +\; cosh\; r\;\; \pr {\phi }_R)\; -\;
{\ca }^L_{\bar z}{\ca }^R_z\; cosh\; r\;].
\eea

\noindent For comparison, we again write the $GWZW$ action, eq.(4), for the
vector gauging in a similar form $[10]$ :
\bea
S^V=&&S^{WZW}\; +\; {k\over 4\pi } \int d^2z\; [\; {\ca }(\bar\pr
{\phi }_R\; +\; cosh\; r\;\; \bar\pr {\phi }_L)\nonumber\\
&& +\;\bar {\ca }(\pr {\phi }_L\; +
\; cosh\; r\;\; \pr {\phi }_R)\; -\;\bar {\ca }{\ca }\; (cosh\; r\;
+\; 1) ]
\eea
It is seen that
the coefficients of the last term in the RHS. of eqs.(7) and (8) are
different. As pointed out earlier, the difference is due to different
gauge choices in the two cases. Now, as in ref.$[11]$, we decouple
the gauge fields by a field redefinition :
\bea
&&{\ca }^R_z\;\equiv\; {\hat{\ca }}^R_z\; +\; {{(\pr {\phi }_L\; +\; cosh\; r
\;\pr {\phi }_R)}\over {cosh\; r}}\nonumber\\
&& {\ca }^L_{\bar z}\;\equiv {\hat{\ca }}^L_{\bar z}\; +\;
{({\bar\pr {\phi }_R\; +\; cosh\; r\;\bar\pr {\phi }_L)}\over
{cosh\; r}}
\eea
and obtain the action for the $CGWZW$ model as,
\bea
S=&&{k\over 4\pi } \int d^2z\; ({1\over 2}{\pr }{\phi
}_L{\bar\pr }{\phi }_L\; +\; {1\over 2}\pr {\phi
}_R{\bar\pr }{\phi }_R\;\nonumber\\
&&+\; {1\over 2}\pr r\bar\pr
r\; +\; {1\over {cosh\; r}}\pr {\phi }_L \bar\pr {\phi }_R
\; -\; {\hat{\ca }}^L_{\bar z}{\hat{\ca }}^R_z\; cosh\; r\;).
\eea
By comparing this action with the following one, describing the
string propagation :
\be
S={1\over {2\pi }}\int\;d^2z\;\big [\big (G_{\mu\nu }+B_{\mu\nu }\big )\pr
x^{\mu }\;\bar\pr x^{\nu }\;\big ],
\ee
one obtains the background metric ${(G_{\mu\nu })}$ and
antisymmetric tensor ${(B_{\mu\nu })}$. For example from eqs.(10)
and (11), by defining an invariant distance :
\be
ds^2\; =\;\Big (-{k\over 4}\;\;\Big )\Big [-dr^2-d{\p _L}^2- d{\p _R}^2 -
{2\over {cosh\; r}}d{\p _L}d{\p _R}\Big ],
\ee
we get
\be
G\; =\;\left (\matrix {{-1} & {0} & {0}\cr {0} & {-1} & {-{1\over
{cosh\; r}}}\cr {0} & {-{1\over {cosh\; r}}} & {-1}\cr }\right ),
\ee
and similarly
\be
B\; =\; \left (\matrix {{0} & {0} & {0}\cr {0} &{0} & {-{1\over
{cosh\; r}}}\cr {0} & {{1\over {cosh\; r}}} & {0}\cr }\right ).
\ee
As in ref.$[1]$, given $G_{\mu\nu }$ and $B_{\mu\nu }$, eqs.(13)
and (14), the background value for the dilaton '$\p $'
can be determined by solving the field equations of the string
effective action. This procedure is followed later in the paper. The resulting
dilaton background is,
\be
\p\; =\; -\; ln\; cosh\; r\; +\; const.
\ee
It can also be shown that the background dilaton of eq.(15) can be
generated from eq.(10) by doing the gauge field integration in the
corresponding path integral and by regularizing a
$\; det\;\big (-{{2{\pi }^2}\over {k\; cosh\;
r}}\big )$ factor as in ref.$[7]$.

We now show that $G_{\mu\nu }$, $B_{\mu\nu }$ and $\p $ fields
obtained above are in fact the solutions of the string effective action.
A characteristic feature of the backgrounds in eqs.(13) -
(15) is that they are dependent only on a single coordinate. String
effective action with such background fields has been studied
extensively by Meissner and Veneziano $[12]$. They worked with time dependent
backgrounds, however their formalism can be applied
to our case by replacing $t$ with $r$ in their paper.
Following ref.$[12]$, it can be shown that for the backgrounds of the
type discussed
in the previous paragraph, $G$ and $B$ can alaways be brought to the form,
\be
G\;\equiv\;\Bigg (\matrix {{-1} & {0}\cr {0} & {\cg }\cr }\Bigg )\;\;\;
;\;\;\;\;\; B\;\equiv\; \Bigg (\matrix {{0} & {0}\cr {0} & {\cb }\cr }
\Bigg )
\ee
where the first entry in a row or column of matrices $G$ and $B$ in
eq.(16) denotes the $r$ coordinate. To the lowest order in
${\a }^{'}$, the string tension, the complete set of field equations
following from the string effective action for this case are :
\be
(\dot \P )^{2}-{1\over 4}Tr[({\cg }^{-1} \dot {\cg} )({\cg }^{-1}
\dot {\cg })]+ {1\over 4}Tr[({\cg }^{-1} \dot {\cb } )({\cg }^{-1}
\dot {\cb })]-V =0\; ,
\ee
\be
({\dot {\P }})^{2}-2{\ddot {\P }}+{1\over 4}Tr[({{\cg }^{-1}
{\dot {\cg }}})({\cg }^{-1}{\dot {\cg }})]- {1\over 4}Tr[({{\cg }^{-1}
{\dot {\cb }}})({\cg }^{-1}{\dot {\cb }})]-V + {{\pr V}
\over{\pr {\P }}}=0\; ,
\ee
\be
-\dot {\P }{\dot {\cg }}+{\ddot {\cg }}-{\dot {\cg}}{\cg }^{-1}
{\dot {\cg }}- {\dot {\cb}}{\cg }^{-1}{\dot {\cb  }}=0,
\ee
\noindent and
\be
-\dot {\P }{\dot {\cb }}+{\ddot {\cb }}-{\dot {\cb}}{\cg }^{-1}
{\dot {\cg }}- {\dot {\cg}}{\cg }^{-1}{\dot {\cb }}=0,
\ee
where
\be
\P\;\equiv\;\p\; -\; ln\;\sqrt{det \cg}
\ee
and dots denote the derivative with respect to the $r$ coordinate.
By comparing eqs.(13) and (14) with eq.(16) we have, for our solutions,
\be
\cg\; =\;-\;\Bigg (\matrix {{1} & {{1\over {cosh\; r}}}\cr
{{1\over {cosh\; r}}} & {1}\cr }\Bigg )\;\; ,\;\;\;\;\cb\;=\; -\;
\Bigg (\matrix {{0} &
{{1\over {cosh\; r}}}\cr {-{1\over {cosh\; r}}} & {0}\cr }\Bigg ).
\ee
This implies,
\be
{\cg }^{-1}\; =\; -\; {coth^2r}\;\Bigg (\matrix {{1} &
{-{1\over {cosh\; r}}}\cr {-{1\over {cosh\; r}}} & {1}\cr }\Bigg ),
\ee
\be
Tr\; ({\cg }^{-1}\dot {\cg })^2\; =\; 2\;\Big ({1\over {sinh^2r\;
cosh^2r}}\; +\; {1\over {sinh^2r}}\Big ) ,
\ee
\be
Tr\; ({\cg }^{-1}\dot {\cb })^2\; =\; 2\;\Big ({1\over {sinh^2r\;
cosh^2r}}\; -\; {1\over {sinh^2r}}\Big )
\ee
and
\be
\dot\P\;\equiv\; \dot\p\; -\; {1\over 2}{\Big ({{det\;\dot {\cg }}\over
{\det \; {\cg }}}\Big )}\; =\; -\; {coth\; r}.
\ee
Using eqs.(24) - (26) it is observed that the field eq.(17)
is satisfied for a constant potential $V=1$. It can be checked that
this is the same value as
for the ungauged $WZW$ model. Also, since $\ddot {\P }\; =\; {1\over
{sinh^2r}}$ , eq.(18) is satisfied for the same value
$V=1$. Eqs.(19) and (20) can be satisfied by using the following
expressions :
\be
-\dot\P\dot {\cg }=\; {1\over {cosh\; r}}\Bigg (\matrix {{0} &
{1}\cr {1} & {0}\cr }\Bigg )\;\; ,\;\;\;\;
-\dot\P\dot B =\; {1\over {cosh\; r}}\Bigg (\matrix {{0} &
{1}\cr {-1} & {0}\cr }\Bigg )\; ,
\ee
\be
\ddot {\cg } =\;\left ( {2\over {cosh^3r}} - {{1\over {cosh\;
r}}}\right )
\Bigg (\matrix {{0} & {1}\cr {1} & {0}\cr }\Bigg )\; ,\;\;\;
\ee
\be
\ddot {\cb } =\;\left ( {2\over {cosh^3r}} - {{1\over {cosh\;
r}}}\right )
\Bigg (\matrix {{0} & {1}\cr {-1} & {0}\cr }\Bigg )\; ,
\ee
\be
\dot {\cg }{\cg }^{-1}\dot {\cg } =\; {1\over {cosh^3r}}\;
\Bigg (\matrix {{-cosh\; r} & {1}\cr {1} & {-cosh\; r}\cr }\Bigg )\;\; ,
\ee
\be
\dot {\cb }{\cg }^{-1}\dot {\cb } =\; {1\over {cosh^3r}}\;
\Bigg (\matrix {{cosh\; r}
& {1}\cr {1} & {cosh\; r}\cr }\Bigg )
\ee
and
\be
\dot {\cb }{\cg }^{-1}\dot {\cg } =\; {1\over {cosh^3r}}\;
\Bigg (\matrix {{-cosh\; r}
& {1}\cr {-1} & {cosh\; r}\cr }\Bigg )
\;=\; -\;\Big (\dot {\cg }{\cg }^{-1}\dot {\cb }\Big )^T\; .
\ee

Therefore we have explicitly shown that the solution, eqs.(13) - (15),
satisfy the field equations of the three dimensional string
effective action. The scalar curvature $R$ can be computed as $[12]$,
\be
R\;  =\; 2\;\big [{{\pr }^2\over {{\pr }r^2}}
(ln\;\sqrt {det\;\cg })\big ]\; +\; {\big [{{\pr }\over {{\pr }r}}
(ln\;\sqrt {det\;\cg })\big ]^2}\; +\; {1\over 4}\; Tr\; ({\cg
}^{-1}\dot {\cg })^2
\ee
and we find for our case $\; R = -\;{7\over 2}\; {1\over {cosh^2r}}$.
Therefore the target space has curvature singularity. However
it does not occur for any real value of $r$ in the coordinate system
of our choice.

Next we show that the solutions, eqs.(13) - (15), can be transformed to a
three dimensional uncharged black string
by an $O(2,2)$ transformation acting on $\cg $, $\cb $
and $\P $.

It was demonstrated in ref.$[12]$ that the string
effective action, as well as the equations of motion, (17) - (20), in
space-time
dimension $D\;(= d +1)$ have an invariance under an $O(d,d)$
transformation acting on a matrix,
\be
M\;\equiv\;\Bigg (\matrix {{{\cg }^{-1}} &
{-{\cg }^{-1}\cb }\cr {{\cb }{\cg }^{-1}} & {{\cg } - {\cb }{\cg
}^{-1}{\cb }}\cr }\Bigg )
\ee
as $M\rightarrow\tilde M =\o\; M\;\o ^T $ and
$\P\;\rightarrow\;\P $, where
$\o $ is an $O(d,d)$ matrix satisfying
$\o^T\;\eta\;\o =\eta $ and $\eta\equiv\Big (\matrix {{0} &
{I}\cr {I} & {0}\cr }\Big )$. Therefore, new solutions ${({\tilde M})}$ can be
generated from the old ones (M) through this transformation. To show
the relation to the uncharged black string, we apply this
procedure to our case. We find that for $\cg $
and $\cb $ as in eqs.(22 ) and for an $O(2,2)$ matrix $\o $:
\be
\o\;\equiv\;\left (\matrix {{1\over {\sqrt 2}} & {0} & {0}
& {-{1\over {\sqrt 2}}}\cr {0} & {1\over {\sqrt 2}} & {1\over {\sqrt
2}} & {0}\cr {0} & {-{1\over {\sqrt 2}}} & {1\over {\sqrt 2}} &
{0}\cr {1\over {\sqrt 2}} &{0} & {0} & {1\over {\sqrt 2}}\cr }\right )
\ee
we have,
\be
{\tilde M}\;=\;\Bigg (\matrix {{{\tilde {\cg }}^{-1}} & {0}\cr {0} &
{\tilde {\cg }}\cr }\Bigg )
\ee
where
\be
\tilde {\cg}\;=\; -\; {(cosh\; r\; +\; 1)\over {sinh^2 r}}
\Bigg (\matrix {{cosh\; r} & {1}\cr {1} &
{cosh\; r }\cr }\Bigg )\; .
\ee

By comparing eqs.(34) and (36)  it is noticed that in the $O(2,2)$
transformed coordinate system, the
antisymmetric tensor $\cb $ is absent. Since $\P $ remains unchanged
under this transformation, therefore we get
$$\dot {\tilde {\p}} - {1\over 2}{\Big ({{det\;\dot {\tilde
{\cg }}}\over {\det \; {\tilde {\cg }}}}\Big )}\;=\; -\; {coth\; r}$$
which implies, using $(det\; {\tilde{\cg }})=coth^2{r\over 2}$,
\be
\tilde {\p}\; =\; -ln\; sinh^2\Big ({r\over 2}\Big )\; +\; const.
\ee
To show that $\tilde {\cg }$ and $\tilde{\p} $ in eqs.(37) and (38) are in
fact the three dimensional uncharged black string, we now diagonalize
$\tilde {\cg }$ by an orthogonal transformation :
\be
{\tilde {\cg }}^{'} \equiv\; O\; {\tilde {\cg }}\; O^T =\;\Bigg (\matrix
{{-coth^2{({r\over 2})}} & {0} \cr {0} & {-1} \cr }\Bigg )
\ee
where $\; O = {1\over {\sqrt 2}}\Big
(\matrix {{1} & {1}\cr {1} & {-1}\cr }\Big ).$

 From the above results we see that
the three dimensional metric $G$ of eq.(13) is transformed,
under combined $O(2,2)$ ${\;(\o )\; }$ and orthogonal
transformations ${(O)}$, to
\be
{\tilde G}^{'}\; =\;\left (\matrix {{-1} & {0} & {0}\cr {0} &
{-coth^2{({r\over 2})}} & {0}\cr {0} & {0} & {-1}\cr }\right )
\ee
The invariant distance corresponding to the metric ${\tilde G}^{'}$
is therefore,
\be
ds^2=\; \Big ({k\over 4}\Big )\;\Big [ dr^2\; +\; coth^2{({r\over 2})}\;
dx^2\; +\; dy^2\Big ]
\ee
where $x$ and $y$ are the corresponding transformed coordinates.

It is now recognised that the metric in eq.(41) represents a target
space which is the cross product of the Euclidean two dimensional "dual"
black hole $[1,10]$ with a flat one dimensional
coordinate. The solution for the
dilaton in eq.(38) is also the one corresponding to the "dual" two dimensional
black hole. From the three dimensional point of view, as pointed out in
ref.$[4]$ , this solution can be interpreted as the uncharged "dual"
black string solution.

In fact our original solution, eqs.(13) and (14), for $G$ and $B$, is also
related to the charged black string even prior to
the $O(2,2)$ transformation. If we just diagonalize the $G$ in eq.(13)
we get the invariant distance :
\be
ds^2=\;\Big ({k\over 4}\Big )\Big [\; dr^2\; +\; \big ({1\over {cosh\; r}}\;
+\; 1\big )\; dx^2\; +\;\big ({1\over {cosh\; r}}\; -\; 1\big ) dy^2\Big ]
\ee
and the antisymmetric tensor :
\be
B\;=\; {1\over {cosh\; r}}\Bigg (\matrix {{0} & {0} & {0}\cr {0} & {0}
&{1}\cr {0} & {-1} & {0}\cr }\Bigg )\; .
\ee
Then by defining $\;\rho = cosh\; r\; $ and $\; y = i\; t$
we have,
\be
ds^2=\Big ({k\over 4}\Big )\Big [-\; \Big (1\; -\; {1\over {\rho}}\Big )dt^2\;
+\;\Big (1\; +\;
{1\over {\rho }}\Big )dx^2\; +\; \Big (1\; -\; {1\over {\rho
}}\Big )^{-1}\Big (1\; +\; {1\over {\rho }}\Big )^{-1}{d{\rho
}^2\over {{\rho }^2}}\Big ],
\ee
\be
B\; =\; {i\over {\rho }}\Bigg (\matrix {{0} & {0} & {0}\cr {0} & {0}
&{1}\cr {0} & {-1} & {0}\cr }\Bigg )
\ee
and
\be
\p\; =\; -\; ln\;\rho\; +\; const.
\ee
This solutions matches with the one in ref.$[4]$ for the charged
black string if one takes $\;\lambda =-{1\over 2}$ and defines ${\;\rho
}=2\;\hat r$ in eqs.(12) and (13) of this reference.
However the $B$ field in this case is imaginary. This
field can be made real by keeping $t$ space-like.
The value $\;\lambda =-{1\over 2}$ in ref.$[4]$ implies that the
free boson '$x$' of the $\; [SL(2,R)\times R]\; $  $WZW$ model is time
like.

To conclude, in this paper the background graviton, dilaton and
antisymmetric tensor fields resulting from the chiral $U(1)$ gauging
of the $SL(2,R)$ $WZW$ model were obtained. It was shown that the
backgrounds satisfy the field equations of the string effective
action and target space has curvature singularity. Relation of our
solution with the three dimensional black string has been shown.
An interesting aspect of
the chiral gauging is that the resulting model has
the same target space dimension as the original one. Whereas due to the
vector or axial gauging the target space dimension is reduced. It
will be interesting to examine the duality properties of the $CGWZW$
models as in the vector and axial cases. Also, since the chiral gauging is
left-right assymmetric in nature, it should be possible to
obtain exact conformal field theory solutions for the heterotic strings
in nontrivial background using $CGWZW$ models.

\vfil
\eject

\baselineskip 12pt
\vfil
\eject

\end{document}